\documentstyle[11pt,newpasp,twoside,epsf]{article}
\begin{document}
\title{The Nuclei of Nearby Radio-Loud Ellipticals}
 \author{G. A. Verdoes Kleijn, P. T. de Zeeuw }
\affil{Leiden Observatory, Postbus 9513, 2300 RA, Leiden, The Netherlands}
\author{S. A. Baum, C. P. O'Dea, R. P. van der Marel, C. Xu}
\affil{Space Telescope Science Institute, 3700 San Martin Drive, Baltimore, MD 21218, USA}
\author{C. M. Carollo, J. Noel-Storr}
\affil{Columbia University, Dept.~of Astronomy, New York, NY 10027, USA}
\begin{abstract}
We have observed a complete sample of 21 nearby ($D < 70h^{-1} {\rm
Mpc}$) Fanaroff \& Riley Type~I galaxies with {\tt HST/WFPC2} and
detected dust disks and lanes in 19 of them. The radio jets are
roughly perpendicular to the dust which is used to constrain the
Doppler boosting factors of the radio jet and cores. The VLBA core
flux correlates with the central H$\alpha$+[NII] flux which
might indicate that the VLBA core is dominated by an isotropic
component. Twelve galaxies show nuclear optical sources. We discuss
various possible origins for this emission.
\end{abstract}

\section{Introduction}
The Fanaroff \& Riley Type I (FR I) radio galaxies in the nearby
($z<0.03$) universe can be characterized as early-type galaxies with
jets emanating from an AGN which is powered by a black hole
(BH). These AGN display only weak nuclear optical line and continuum
emission. The FR I stellar hosts and their unresolved cores bear
resemblance to both normal early-type galaxies with radio cores, which
constitute a considerable fraction of the nearby early-type population
($\sim 50\%$ at $M_B=-22$, cf.~Sadler 1997) and to early-type galaxies
with unresolved blue optical spikes (Carollo et al.~1997). On the
other hand, the FR I galaxies appear to be scaled-down versions of
powerful radio galaxies and quasars in the distant universe which have
strong nuclear optical line and continuum emission. For a physical
understanding of the connection between active and normal galaxies, it
is important to determine how these low-luminosity active nuclei and
their jets form and evolve. Relevant questions are then: where did the
accreted matter come from, how does this accretion trigger jet
formation, and what is the origin of the optical AGN luminosity? We
are using {\tt HST} to study the centers of a complete sample of FR I
radio galaxies and to isolate weak optical nuclear activity from the
stellar background (Verdoes Kleijn et al.~1999). Here we discuss
further interpretation of the {\tt WFPC2} data in combination with
VLBA radio data presented by Xu et al.~(2000).

\section{Orientation of Dust and Radio Jets}
We detected dust in the centres of 19 galaxies (Fig.~1). The dust
extent ranges from a few hundred pc to a few kpc. In eleven galaxies,
the dust morphology is smooth and elliptical (a `disk'), while in
eight galaxies it is filamentary with wisps and bends (a
`lane'). Lanes are roughly perpendicular to the radio jets: the
position angle difference $\Delta {\rm PA}$ is in the range $68\deg -
90\deg$. Processes that cause such a preferred orientation are
discussed in e.g., Quillen \& Bower (1999). By contrast, all disks
closely align with the galaxy major axis. The $\Delta {\rm PA}$ of the
disk major axis and radio jet is in the range range $23\deg -
90\deg$. One can assume that (i) dust disks are circular and (ii) the
brightest side of the radio jet is approaching the observer. The
allowed range of the {\sl intrinsic} angle $\alpha$ between the sides
of both the dust disk symmetry axis and the radio jet axis that are
closest to the direction of the line of sight, can then be
computed. By definition, the radio jet inclination is in the range
$0\deg-90\deg$, but $\alpha$ can take a value in the range $0\deg -
180\deg$. The right panel in Fig.~1 shows that the allowed range of
$\alpha$ for each dust disk is constrained such that the dust disk
symmetry axis and the radio jet axis tend to `align' (i.e., $\alpha <
90\deg$).  Further analysis indicates that the angle with the line of
sight for jet and disk are expected to differ typically by only $\sim
10\deg - 20\deg$. A similar result was derived by Capetti \& Celotti
(1999) for a small pilot sample.

\begin{figure}[t!]
\vbox {
  \begin{minipage}[l]{1.0\textwidth}
   \hbox{
       \begin{minipage}[l]{0.3\textwidth}
       {\centering \leavevmode \epsfxsize=\textwidth 
        \epsfbox{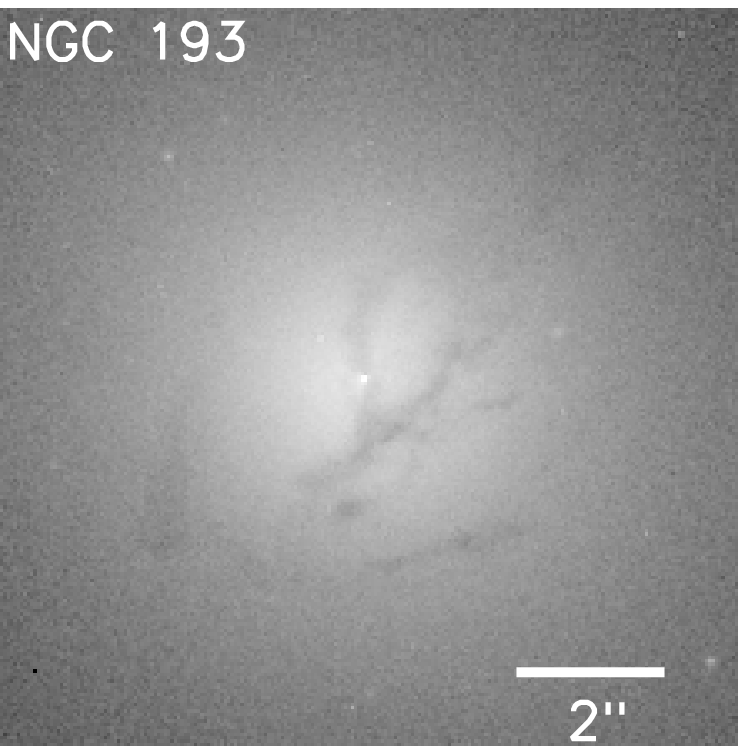}}
       \end{minipage} \  \hfill \
       \begin{minipage}[l]{0.3\textwidth}
       {\centering \leavevmode \epsfxsize=\textwidth 
        \epsfbox{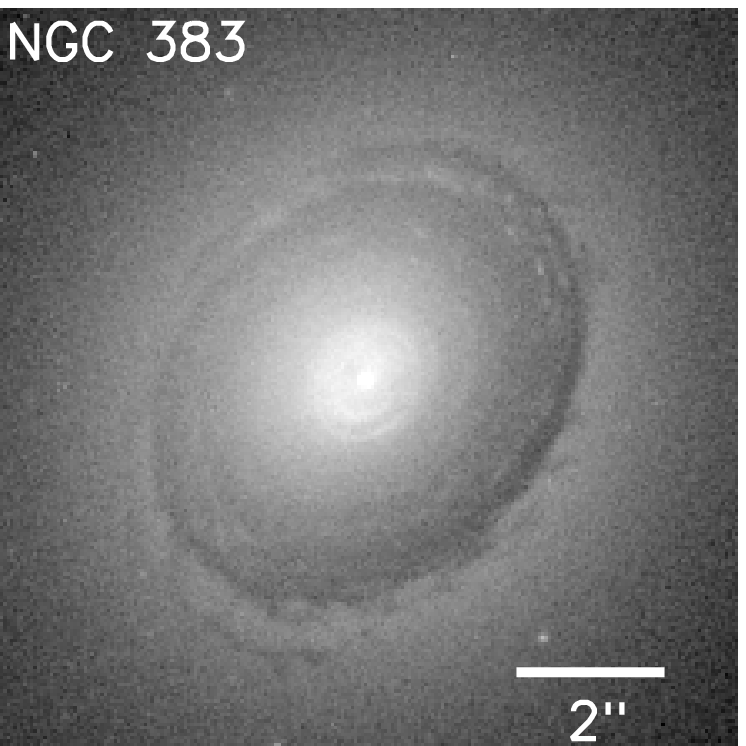}}
       \end{minipage} \  \hfill \
       \begin{minipage}[r]{0.4\textwidth}
       {\centering \leavevmode \epsfxsize=\textwidth 
        \epsfbox{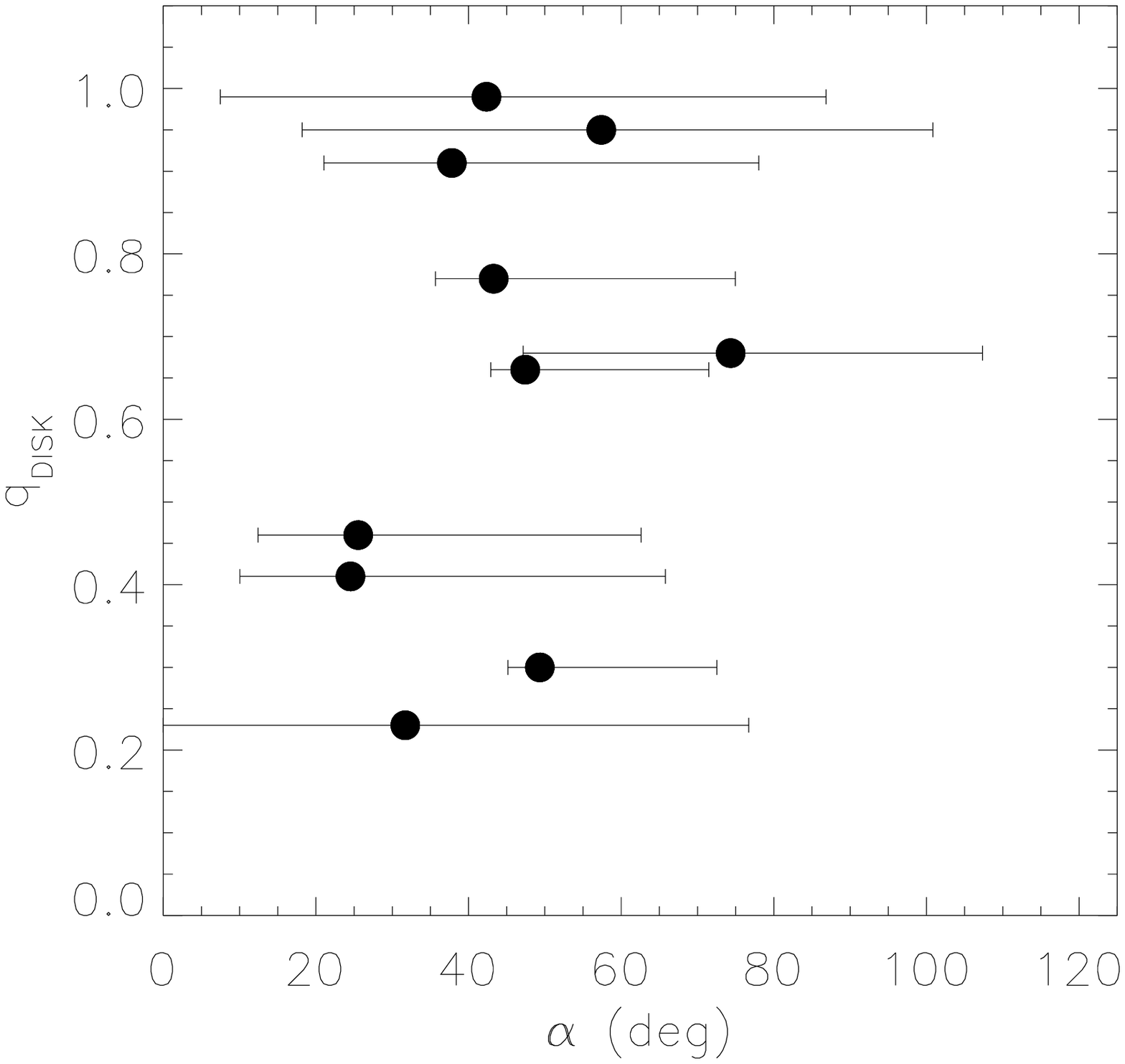}}
       \end{minipage} \  \hfill \
  }
  \end{minipage} \  \hfill \
  \begin{minipage}[l]{1.0\textwidth}
\caption{{\bf Left and middle:} $V$-band images of two sample
galaxies. NGC 193 has a dust lane and NGC 383 has a dust disk. Both
galaxies have a blue nuclear optical source. {\bf Right:} the axis
ratio of the dust disk plotted versus the range of the angle $\alpha$
between the sides of the dust disk symmetry axis and the radio-jet
axis, that are closest to the direction of the line of
sight. Horizontal bars indicate the allowed range for $\alpha$ and the
solid dots indicate the median $\alpha$.}
\end{minipage} }
\vspace*{-0.4cm}
\end{figure}

\section{Doppler Boosting of Radio Jets and Cores}
Xu et al.~(2000) report the surface brightness ratio $S$ of VLBA jet
and counter jet at 1670 MHz for the sample galaxies. If the jets are
intrinsically symmetric and ejected in opposite directions, $S$
depends on jet inclination and velocity (e.g., Urry \& Padovani
1995). As discussed in the previous section, the dust disk inclination
is a reasonable estimator for the jet inclination and can constrain
the jet velocity. Fig.~2 plots $S$ versus jet inclination together
with model predictions for constant jet velocity $v$, assuming an
isotropic continuous jet with no preferred direction of the magnetic
field and a radio spectral index $\alpha=0.75$ ($f_{\nu} \sim
{\nu}^{-\alpha}$) typical for jets. The observed values of $S$ cannot
constrain $v$ very well given that most are lower limits (i.e., the
counter jet is not observed) and given the $\sim 20\deg$ uncertainty
of the jet inclination. For NGC 315 we can constrain the Lorentz
factor to be $\gamma > 10$.

The VLBA radio-core flux (unresolved at the parsec scale) correlates
tightly with central {H$\alpha$+[NII]} flux (Fig.~2). Interestingly, a
similar correlation is found for radio-core galaxies (Ho 1999). If the
{H$\alpha$+[NII]} flux is emitted isotropically and correlates tightly
with intrinsic VLBA core flux, the observed scatter in the correlation
might be due to Doppler boosting. The low scatter in the correlation
then constrains $\gamma < 2$. Indeed, no dependence of the VLBA core
flux on dust disk inclination is found. Thus it seems that the VLBA
core flux is dominated by a relatively isotropic component instead of
a relativistic jet. This isotropic component might have uncollimated
relativistic motion.

\begin{figure}
\plotone{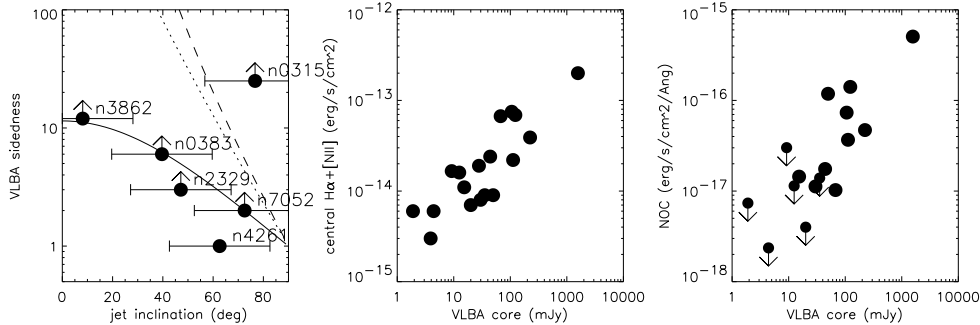}
\caption{{\bf Left:} the VLBA jet to counter jet surface brightness
ratio $S$ versus dust disk inclination. The arrows indicate lower
limits. The curves describe $S$ for an isotropic continous jet with no
preferred direction of the magnetic field, a radio spectral index
$\alpha=0.75$ and $\gamma = 1.1$ (solid line), $\gamma = 2.0$ (dotted
line) and $\gamma = 10.0$ (dashed line). The horizontal error bars
indicate the uncertainty in the jet inclination. {\bf Middle:}
{H$\alpha$+[NII]} flux in the central $1''$ versus VLBA core flux
(1670 MHz). {\bf Right:} flux of the nuclear optical sources
($I$-band) detections and upper limits versus VLBA core flux.}
\end{figure}

\section{Nuclear Optical Sources}
Twelve galaxies show blue nuclear optical sources (NOS) unresolved
with {\tt WFPC2}, corresponding to sizes of tens of parsec or
less. The right panel in Fig.~2 shows NOS flux versus VLBA radio core
flux. The observed correlation is significant at the 99.999\% level
(generalized Kendall's Tau test). This agrees with results by
Chiaberge, Capetti, \& Celotti (1999). Radio core emission is
generally assumed to be self-absorped synchrotron emission. The
correlation might indicate that the NOS is also synchrotron
emission. The radio-to-optical spectral index varies between 0.43 and
0.85. These values are consistent with those found for galaxies in our
sample with extended optical jets: 3C 66B, 3C 31 and M87 (Butcher, van
Breugel \& Miley 1980; Biretta, Stern, \& Harris 1991). The slope of
the log-log correlation, $s=1.04 \pm 0.24$, although not well
determined, is consistent with a power-law spectral energy
distribution (SED) from radio to optical wavelengths.\looseness=-2

Alternatively, the NOS might be emission from the accretion disk
and/or flow. For example, Di Matteo et al.~(2000) obtain a reasonable
fit to the nuclear radio to X-ray SED of M87 using an ADAF model with
matter outflow. However, the models that fit the observed X-ray SED
underpredict the nuclear optical emission by a factor of $\sim
4$. Furthermore, Di Matteo et al.~note evidence for a contribution to
the flux by the synchrotron jet at radio and millimeter fluxes. If the
accretion disk is inside an optically thick torus, the NOS detection
rate implies an opening angle $\sim 130\deg$. Broad emission-line
regions are commonly detected in powerful AGN but typically not
detected in FR Is. The large opening angle would then suggest BLRs are
generally not present in FR I galaxies

Finally, the NOS might be produced by a nuclear starburst. The high
NOS detection rate would require a continuous starburst on time scales
on the order of the radio-source lifetime, typically estimated to be
$10^{7-8}$ yr (cf.~Chiaberge, Capetti, \& Celotti 1999). However,
detailed studies of the optical nuclear spectra of M87 and M84
indicate their NOS are indeed not produced by a starburst but by
non-thermal AGN emission ( Kormendy 1992; van der Marel 1994; Bower et
al.~2000). \\

\end{document}